\documentstyle[prb,aps,twocolumn]{revtex}
\large
\begin{document}
\draft
\twocolumn[\hsize\textwidth\columnwidth\hsize\csname
@twocolumnfalse\endcsname

\title{Investigation of thermal and magnetic properties of
defects in a spin-gap compound NaV$_2$O$_5$}

\author{A.~I.~Smirnov, S.~S.~Sosin}
\address{P.~L.~Kapitza Institute for Physical Problems RAS, 117334
Moscow, Russia}

\author{R.~Calemczuk}
\address{Centre d'Etudes Atomique, Grenoble, Cedex 9 France}

\author{ V.~Villar, C.~Paulsen }
\address{Centre Nationale de Recherches Scientifique, Grenoble, Cedex 9
France}

\author{M.~Isobe, Y.~Ueda}
\address{Institute for Solid State Physics, University of Tokyo, 7-22-1
Roppongi, Minato-ku, Tokyo 106, Japan}

\date{\today}
\maketitle

\begin{abstract}
\widetext
\leftskip 54.8pt
\rightskip 54.8pt

The specific heat, magnetic susceptibility and ESR signals
of a Na-deficient vanadate Na$_x$V$_2$O$_5$ ($x=$1.00 - 0.90) were
studied in the temperature range 0.07 - 10 K, well below the
transition point to a spin-gap state. The contribution of defects
provided by sodium vacancies to the specific heat was observed. It
has a low temperature part which does not tend to zero till at
least 0.3 K and a high temperature power-like tail appears
above 2 K. Such dependence may correspond to the existence of
local modes and correlations between defects in V-O layers. The
magnetic measurements and ESR data reveal $S=$1/2 degrees of
freedom for the defects, with their effective number increasing in
temperature and under magnetic field. The latter results in the
nonsaturating magnetization at low temperature. No long-range
magnetic ordering in the system of defects was found. A model for the
defects based on electron jumps near vacancies is proposed to
explain the observed effects. The concept of a frustrated
two-dimensional correlated magnet induced by the defects is considered
to be responsible for the absence of magnetic ordering.

\end{abstract}


]

\narrowtext

\section{Introduction}

The sodium vanadate NaV$_2$O$_5$ is a unique compound with
1D-magnetic structure formed within V-O layers due to the special
arrangement of vanadium-oxygen orbitals. An intensive study of
this compound began when a phase transition into a dimerized
spin state was discovered \cite{Ueda1}. According to the present
model, vanadium-oxygen bonds give rise to a ladder structure with
one electron (and one spin) per rung which should be shared
between two vanadium ions
\cite{Horsch,Smolinski,Schnering,ThalmeierFulde}. Electrons are
localized at rungs due to the Coulomb repulsion in accordance with
the dielectric state of the crystal. In the high temperature phase
the average charge of each vanadium ion is +4.5. The phase
transition into a spin-gap state at $T=$36 K was at first
interpreted as a spin-Peierls transition \cite{Ueda1}. Further
theoretical and experimental investigations
\cite{MostovoyKhomskii,Fukuyama,Smirnov99} point to a charge ordering phase
transition that
results in a zigzag charge distribution along the ladders, with a doubling
of the lattice period and with alternating exchange interactions. The latter
is responsible for the spin-gap opening according to
Ref.~\CITE{Bulaevski}. The spin-gap causes an exponential freezing
of the spin part of the magnetic susceptibility observed below the
transition \cite{VasilevSmirnov}.

There are two reasons why the problem of defects in the spin-gap
crystals is of great interest. First, soliton-type spin
clusters may be formed around imperfect spins or nonmagnetic
impurities \cite{Khomskii,Fukuyama2,Buzdin}. These clusters are
mesoscopic multispin objects with a microscopic total spin. The
multispin nature of such clusters was confirmed by ESR experiments
in the Ni-doped spin-Peierls compound CuGeO$_3$ \cite{Glazkov}.
Second, the defects in a spin-gap (and hence, nonmagnetic) matrix
can induce the long range antiferromagnetic ordering due to
correlation of spins in neighboring clusters.  Such an induced
antiferromagnetic ordering was observed in the dimerized phase of
the spin-Peierls crystal CuGeO$_3$ \cite{Regnault1} and in the
Haldane compound PbNi$_2$V$_2$O$_8$ \cite{Uchiyama}.

The substitution of magnetic ions by other magnetic or nonmagnetic
ions in NaV$_2$O$_5$ is not yet reported. The only reliable method
to embed magnetic defects into this spin-gap crystal is the
creation of sodium vacancies \cite{IsobeUeda_deficit}. Each
vacancy causes a 3d-electron shared by two vanadium ions to couple
with an oxygen ion, thus making a next-to-defect rung of the
vanadium ladder empty ({\it i.e.} nonmagnetic). This kind of
defects should therefore be equivalent to the diamagnetic dilution
or cutting the dimerized $S=$1/2 chains. Nevertheless, one should
note that two ladder rungs situated next to the sodium vacancy are
electrically equivalent which allows the electron to occupy either site or
to jump between the rungs.

The present work is devoted to the study of low temperature
thermal and magnetic properties of Na-deficient NaV$_2$O$_5$
in order to search for long-range magnetic ordering stimulated by
defects. It is also interesting to compare the behavior of jumping
magnetic defects in this kind of a dimerized spin matrix to the
behavior of strongly localized defects in a spin-Peierls compound
CuGeO$_3$. No transition to an ordered phase was found down to a
temperature of 0.07~K, however unusual magnetic and thermal properties
due to the defects were observed and a model of jumping defects to
describe these properties has been developed. The problem of
delocalized magnetic defects is suggested for more detailed theoretical
investigation. Further experiments to clarify the question are
proposed.

\section{Experiment}

\subsection{Samples}

The sodium deficient single crystals were prepared by the techniques
described in Ref.~\CITE{IsobeUeda_deficit}. Single crystals of
the stoichiometric compound were embedded in a large quantity of the
Na-deficient powder and were heated for one week. The resulting
content of sodium in the crystals after heating was controlled
with accuracy of about 1 \% by comparing the results of x-ray and
magnetic measurements to those obtained for reference powder
samples. The content of sodium in the powder was determined from the
molar ratios of the initial reagents.

\subsection{Specific heat}

The specific heat of Na$_x$V$_2$O$_5$ samples with different
sodium deficiencies was measured in the temperature range 0.3 - 8
K. The samples were put onto the holder mounted by thin
thermoinsulating wires inside the massive ring equipped by a
thermometer. The temperature difference between the holder and the
ring was controlled by Au-Fe thermocouple. The current induced by
thermoelectric power of the thermocouple was detected by SQUID.
The signal from the SQUID was used in a feed back circuit to
support the temperature difference equal to zero by heating the
ring. Supplying the known power $P$ to the holder with the sample
we measured their temperature in real time, thus obtaining the
total heat capacity. More detailed description of the installation
and the experimental technique including the correction for the
parasitic power etc. is given in Ref.\CITE{Calem}. The specific
heat of a stoichiometric sample was found to be cubic in
temperature according to previous measurements \cite{HeatCap}. The
molar heat capacities of the imperfect samples appeared to be much
larger in the whole temperature interval. The contribution of the
defects to the heat capacity obtained as a difference between the
total value of the molar specific heat of several Na$_x$V$_2$O$_5$
samples and that of $x$=1.00 sample is shown in Fig.~1.  The
specific heat of a perfect (x=1.00) sample fitted by $T^3$-line is
also given as a reference. The defects contribution may be divided
into two main components:  the low-temperature part remaining
non-zero till at least 0.3 K (see Fig.~2) and the high-temperature
power-like tail. The temperature dependence above 4 K is
approximately quadratic. The low temperature part increases
monotonously in the concentration of defects ({\it i.e.} in $1-x$)
while the value of the quadratic part reaches its maximum at
$1-x\simeq$0.04. A test measurement of the specific heat under
magnetic field was performed above 4 K. It was found to be field
independent up to 7~T with an accuracy of 10\%. Even at the
largest vacancy concentration no sign of a phase transition into
an ordered state was observed, in contrast to similar measurements
on diluted CuGeO$_3$.

The contribution of defects to the entropy was calculated
numerically by taking the integral $E=\int\limits_{0.3}^{T}
\frac{\Delta C_P}{T}dT$. The temperature dependence of this
integral divided by the molar entropy of an ensemble of two-level
systems is shown in Fig.~3. One should note that the total
entropies of the samples with $x=$0.96 and 0.92 are almost the
same while the entropy of the $x=$0.98 sample is twice as small.
This probably indicates the appearance of some collective state in
the system of defects at high deficiency concentrations.

\subsection{Magnetic susceptibility}

In order to look at magnetic properties in the low temperature range,
we performed a series of magnetization measurements using a SQUID
magnetometer equipped with a miniature dilution refrigerator in the
temperature range
0.07 - 1.2 K and in magnetic fields up to 8~T. The temperature
dependencies of the magnetization at $H\simeq$0.05 T were obtained
for samples with $x$=1.00, 0.98 and 0.90. The corresponding
inverse susceptibilities are shown in Fig.~4. The susceptibility
of the stoichiometric sample was found to be Curie-like, with the
effective concentration of free $S=$1/2 spins being equal to
6$\cdot$10$^{-4}$. The small positive Weiss constant
$\theta\simeq$0.03~K
corresponds to weak antiferromagnetic interactions between spins.
The magnetic defects give rise to an additional Curie-like
contribution to the susceptibility which is nevertheless much
smaller than that expected for the system of free $S=1/2$ spins
with the number equivalent to one half of the concentration of
defects. The Weiss constant increases to $\theta\simeq$0.07~K. The
effective concentrations of free $S$=1/2 spins and Curie-Weiss
constants of several samples determined by fitting the magnetic
susceptibility below 1~K to a Curie law are given in Table~1.

Table 1

\vspace{3mm}
\begin{center}
\begin{tabular}{|c|c|c|}
\hline $~1-x~$ &$~\theta$, mK~& Effective
concentration \\ \hline
0&30$\pm$10&6$\cdot$10$^{-4}$\\ \hline
0.02&70$\pm$10&3.7$\cdot$10$^{-3}$\\ \hline
0.1&70$\pm$10&6$\cdot$10$^{-3}$\\ \hline
\end{tabular}
\end{center}

\vspace{3mm}

The inverse susceptibility for samples with small $(1-x)$ values
deviate from a linear temperature dependence for $T>$1~K.
This seems to indicate that the effective concentration of spins
increases with temperature. On the other hand, we cannot rule out that
some of the temperature independent part may be due to non-uniformity
along the sample holder (no corrections have been made to the present
data).

Test magnetization measurements were made for $x=$0.98 sample by
standard Quantum Design SQUID magnetometer in the temperature
range 4~-~77 K. The value of Curie constants obtained from data
fits at 4~-~15 K corresponds to $(1-x)/2$ concentration of defects
which is in agreement with previous results
\cite{IsobeUeda_deficit}.

The field dependence of the magnetization at $T=$0.077~K was also
studied for these samples. Although the temperature dependence of the
susceptibility at low temperature is Curie like, the magnetization
was found to have a component which does not saturate till $H=$8~T
(see Fig.~5). It should be mentioned that at $T=$0.077~K the
saturating field resulting in 90\%-polarization of free $S=1/2$
spins is equal to 0.17~T.

There was no indication of a phase transition down to 0.077~K
in accordance with the specific heat measurements described above.

\subsection{ESR}

The sodium deficiency gives rise to an increase in the ESR
intensity at low temperature. Imperfect samples of NaV$_2$O$_5$
demonstrate the broadening of the ESR line in the temperature
range below 3~K (see Fig.~6) in contrast to the constant value of
the linewidth in stoichiometric samples \cite{VasilevSmirnov}. The
ESR linewidth of magnetic defects caused by Na-vacancies appears
to be narrower than that in a perfect sample which proves the
different nature of the residual defects in a nominally
stoichiometric sample and of the artificial defects in
Na-deficient samples.

\section {Discussion}

As mentioned in the introduction, the sodium deficiency in
NaV$_2$O$_5$ crystals appears to give rise to a new kind of
magnetic defect in spin-gap systems. Unlike the usual nonmagnetic
dilution, when impurity ions are fixed in the basic magnetic
matrix and all the spins are localized, the absence of a sodium
ion leads to the absence of an electron (but not an ion) at a rung
of the vanadium ladder. Because the sodium atoms are positioned
symmetrically in between two rungs (see Fig. 7) one can see that
the resulting unoccupied rung will have two almost equivalent
positions. We suggest that this extra degree of freedom for the
electron position is very important for the system with exchange
interactions and we will try to explain the observed effects
starting from this point. We also assume that the ground state of
the electron is localized at one of the two next-to-vacancy rungs.
In spite of the obvious loss in energy due to electron
localization (of the order of the hopping amplitude \cite{Smolinski}
$t_{\parallel}\sim$0.15 eV),
this assumption seems plausible. Mentioning, that the charge ordering
phase transition observed at $T$=36~K is also followed by localization
of electrons (at the ends of the rungs, with the corresponding
hopping amplitude $t_{\perp}\sim$0.35 eV) one can suggest the
existence of an additional localization mechanism. For example,
the shift of the electron to one of the rungs may break the
symmetry of its potential hole in the vicinity of defect, thus
strongly diminishing the hopping amplitude.

The most obvious consequence of the sodium (electron) deficiency
may be understood neglecting for the exchange alternation. As
mentioned above, the diamagnetic dilution of 1D-magnets usually
cuts magnetic chains into fixed segments with an equal probability
for them to contain an odd or an even number of spins (we shall
refer them to as "odd" and "even" segments). However electron
jumps can result in a disproportion between them. The ground state
energy of a uniform even segment is $E_{even}=-NE_0-a/N$ and
that of an odd one is $E_{odd}=-NE_0+b/N$,
where $E_0\simeq 0.886J$,  $b\simeq 2a\simeq 3.5J$ and $N$ is the
number of spins in a segment \cite{Bonner-Fisher}. Thus, the even
states correspond to a gain in exchange energy. A one electron jump
between "odd-odd" and "even-even" states of two neighboring segments
can change the energy by approximately $\Delta E\sim (a+b)/N$.
Naturally, for long chains this
effect should be strongly damped by dimerization which, nevertheless
gives an additional gain in energy corresponding roughly to the
spin gap energy, which comes from the recovering of "dimers" in
even segments. It is also reasonable to suggest the existence of a
random electrostatic potential in a crystal with defects which
makes the neighboring rungs slightly inequivalent. For simplicity
one can assume it to be constant but of a random sign $U=\pm
\epsilon$. Taking the statistical distribution of segment length
$N$, $f(N)=(1-x)x^N$ ($1-x$ - concentration of chains breaks) we obtain
a set of two-level systems with the following molar heat capacity:

\begin{equation}
C=\frac{1}{2}(1-x)N_A\sum f(N)\left (C(\Delta^-_N)+
C(\Delta^+_N)\right ),
\end{equation}

where $C(\Delta^{\pm}_N)$ is the heat capacity of a two-level system
with the gap \\ $\Delta^{\pm}_N=\mid\Delta E(N)\pm\epsilon\mid$. This
formula describes qualitatively the low temperature part of the
experimental curves for various concentrations of defects (see
Fig.~1,2).

Consider also the behavior of such a system under magnetic field
which splits the energy levels corresponding to the odd states of
the chains. One can easily obtain the following formula for the
magnetization:

\begin{equation}
M(H,T)=\frac{1}{2}(1-x)\mu N_A\sum f(N)\frac{e^{\frac{\mu
H}{kT}}-e^{-\frac{\mu H}{kT}}}{e^{\frac{\mu H}{kT}}+e^{-\frac{\mu
H}{kT}}+e^{\frac{\Delta E(N)\pm \epsilon}{kT}}},
\end{equation}

where the summation is performed over $N$ for $+\epsilon$ and
$-\epsilon$. The result is in surprisingly good agreement with the
corresponding experimental data (see Fig~4,5). This effect has a simple
interpretation: due to a gain in exchange energy, the number of even
segments is initially larger than that of odd ones which is the reason
for the system of defects to be almost nonmagnetic in the ground state.
The effective number of magnetic objects (odd chains) may increase
either by thermal activation or under magnetic field due to a gain in
Zeeman energy.  (Note, that the small steps at theoretical curves on
Fig.~5 result from the discrete distribution of two-level systems by
energies $\Delta E(N)$).  One must emphasize that this very rough model
does not cover the effects of dimerization and correlations between
defects and ladders. In addition it implies that the electrons are
strongly localized at one of two next-to-vacancy rungs. The aim of this
model is only to demonstrate the main features of the thermodynamic
behavior of jumping defects.

The observed low temperature widening of the ESR line in Na-deficient
samples may result from the freezing out of electrons jumps in the
vicinity of defects. When jumps are activated the ESR line should be
narrowed by this kind of motion similar to the magnetic resonance
line in a liquid.

The unexpected result of the specific heat measurements is the
observation of a power-like contribution at relatively  high
temperatures. While the low temperature part can be described in
terms of the thermal activation of local modes (electron jumps)
the power-like tail could indicate correlations of some kind
between defects which may be of at least two origins.

\begin{enumerate}
\item[a).] The correlation between edge spins in segments due to a gain
in exchange energy for "even" chains: The jump of one edge
electron of a segment affects the electron at the other edge
forcing it to shift in the same direction (for even number of
rungs between the vacancies) or in the opposite direction (for odd
number) in order to recover the initial parity (see Fig. 7). Thus,
the initial jump appears to be a perturbation transmitting along
the ladder. These shifts also disturb the zigzag pattern in
neighboring ladders because of the correlations between them,
which allows this perturbation to transmit also in the transverse
direction. One should mention that Fig. 7 shows the perturbations
damaging the dimerized state of the segments. It is also possible
to imagine a situation where the dimerization is restored in
which case it is obviously not the same gain in exchange energy.
A detailed theoretical study of the correlations between such
defects is required.

\item[b.)] The spin clusters arising around the defect in a dimerized
spin-gap matrix: Several spins should be correlated
antiferromagnetically in the vicinity of each defect and the
interchain interaction may cause the long-range AFM ordering
\cite{Khomskii,Fukuyama2,Buzdin}. Such an ordering forces each pair of
spins separated by an empty site to be parallel, otherwise the
interchain interaction appears to be frustrated (see Fig~8). In case of
NaV$_2$O$_5$ this frustration may be caused by AFM exchange $J^*$
between edge spins appearing due to electron jumps. It is probably the
reason for the absence of the spin vacancy induced long range ordering
confirmed in the present work down to $T=$0.077~K.  Nevertheless, the
defects should induce a frustrated 2D-magnet instead of an
antiferromagnet.  This is a strongly correlated system (see, for
example, Ref.~ \CITE{Ramirez}) whose specific heat is known to be
power-like in temperature as observed in our experiment.
\end{enumerate}

It would be interesting to search for the phase transition into an
ordered state in crystals of NaV$_2$O$_5$ with a partial
substitution of V-ions for other ions ({\it e.g} for Ti) fixed in
the lattice. First, the susceptibility resulting from such a
substitution would probably be much larger than those reported in
this work. It may also result in the creation of either an ordered
state or a frustrated magnet, or a spin glass state. It will thus be very
interesting to compare the properties of fixed and jumping
defects and their influence on the spin-gap phase of NaV$_2$O$_5$.
From the theoretical point of view it is very interesting to study the
problem of localization of electrons in the vicinity of
unoccupied rungs of the spin ladder (with and without
dimerization). The problem of the interaction of jumping electrons
with the spin gap matrix has not been studied either.

\section{Conclusions}

The contribution to the specific heat, magnetic susceptibility
and ESR signal provided by defects in the spin gap matrix of
NaV$_2$O$_5$ was observed. The data may be qualitatively described
considering the defects as unoccupied rungs of a ladder structure.
These defects provide spin and jumping degrees of freedom. The
temperature dependence of the specific heat reveals local modes
and correlation between defects. We ascribe local modes to
electron jumps in the vicinity of unoccupied rungs of the ladder
structure. The correlation of defects is explained qualitatively
by a model of perturbations caused by electron jumps transferred
along the ladder and to the neighboring ladders. Another possible
mechanism of the correlation is the generation of magnetic
clusters around defects which result in a frustrated 2D-magnet.
The absence of an induced long range antiferromagnetic order
(in contrast to other spin-gap systems) is interpreted as the
consequence of the geometric frustration of the exchange
interaction between antiferromagnetic clusters. Magnetic
properties of the system of defects are characterized by the
temperature dependent effective spin concentration and by the
nonsaturating magnetization curves. These properties are also described
in the frame of the jumping defect model.

\section {Acknowledgments}

This work was supported by CIES (France), the Russian Fund for
Basic Researches grant 98-02-16572 and INTAS grant  99-0155.
Authors thank J.~Flouquet, D.~I.~Khomskii, M.~V.~Mostovoy,
A.~N.~Vasil'ev and M.~E.~Zhitomirskii for valuable discussions.

{\bf Figure captions}

Fig. 1. The contribution to the specific heat from defects at
various concentrations. Solid line is the quadratic high
temperature approximation, dotted and dashed-dotted lines are
calculations by formula (1) with the parameters $\epsilon$=7 K,
$a+b$=450 K and 350 K for $x$=0.96 and 0.98 respectively, dashed
line is $T^3$-fit to the phonon part of the specific heat (perfect
sample) subtracted from all the data.

Fig. 2. The residual low temperature part of the contribution to
the specific heat from defects. Lines are the same as in Fig. 1.

Fig. 3. The temperature dependence of the molar entropy of the defects.

Fig. 4. The temperature dependence of the inverse susceptibility
for samples with various Na-deficiency, ${\bf H}\parallel b$.
Solid lines are theoretical fits by formula (2) with the
parameters $\epsilon$=8 K, $a+b$=400 K and 300 K for $x$=0.98 and
0.90 respectively. Dashed line is the linear fit of low
temperature points.

Fig. 5. The field dependence of magnetization at $T=$0.077~K,
${\bf H}\parallel b$. Solid lines are theoretical fits by formula
(2) with the parameters as in Fig. 4. Dashed line is the
magnetization of a paramagnet with the effective spin
concentration $(1-x)/2$ (x=0.98).

Fig. 6. The temperature dependence of the ESR linewidth at $f$=36
GHz, ${\bf H}\parallel b$; solid lines are guide-to-eyes.

Fig. 7. Schematic representation of the charge ordered ladder
structure in the ($ab$)-plane of the NaV$_2$O$_5$ crystal
according to Ref.\CITE{MostovoyKhomskii}. Solid lines represent
the paths of exchange interaction. Vanadium ions are positioned at
the intersections of ladder rungs and legs; closed circles are
V$^{4+}$ ions, open circles are V$^{5+}$ ions, grey circles are
ions with jumping electrons; triangles are the projections of
sodium vacancies onto the ($ab$)-plane. The pairs within dashed
ovals are coupled by larger exchange interaction ("dimers"). The
arrows show the direction of correlated electron jumps excited at
finite temperature.

Fig. 8. An antiferromagnetic interchain frustration due to
electron jumps (frustrated region is outlined by dashed
rectangle). For simplicity the zig-zag patterns are shown as
chains.

\end{document}